\documentclass[sigconf]{acmart}
\usepackage{tabularx}
\usepackage{tipa}

\AtBeginDocument{%
  }

\copyrightyear{2025}
\acmYear{2025}
\setcopyright{acmlicensed}
\acmConference[MM '25]{Proceedings of the 33rd ACM International Conference on Multimedia}{October 27--31, 2025}{Dublin, Ireland}
\acmBooktitle{Proceedings of the 33rd ACM International Conference on Multimedia (MM '25), October 27--31, 2025, Dublin, Ireland}
\acmDOI{10.1145/3746027.3758210}
\acmISBN{979-8-4007-2035-2/2025/10}

\begin{document}

\title{Multi-Accent Mandarin Dry-Vocal Singing Dataset: Benchmark for Singing Accent Recognition}










\author{Zihao Wang}
\affiliation{%
  \institution{Zhejiang University} 
  \city{Hangzhou}
  \country{China}
}
\affiliation{%
  \institution{Carnegie Mellon University} 
  \city{Pittsburgh}
  \country{United States}
}
\email{carlwang@zju.edu.cn}

\author{Shulei Ji}
\affiliation{%
  \institution{Zhejiang University} 
  \city{Hangzhou}
  \country{China}
}
\affiliation{%
  \institution{Innovation Center of Yangtze River Delta, Zhejiang University} 
  \city{Hangzhou}
  \country{China}
}
\email{shuleiji@zju.edu.cn}

\author{Le Ma}
\affiliation{%
  \institution{Zhejiang University} 
  \city{Hangzhou}
  \country{China}
}
\email{maller@zju.edu.cn}

\author{Yuhang Jin}
\affiliation{%
  \institution{Zhejiang University} 
  \city{Hangzhou}
  \country{China}
}
\email{3210103422@zju.edu.cn}

\author{Shun Lei}
\affiliation{%
  \institution{Tsinghua University} 
  \city{Shenzhen}
  \country{China}
}
\email{leis21@mails.tsinghua.edu.cn}

\author{Jianyi Chen}
\affiliation{
  \institution{Hong Kong University of Science and Technology} 
  \city{Hongkong}
  \country{China}
}
\email{jchenil@connect.ust.hk}

\author{Haoying Fu}
\affiliation{
  \institution{Mei KTV} 
  \city{Beijing}
  \country{China}
}
\email{326452438@qq.com}

\author{Roger B. Dannenberg}
\affiliation{
  \institution{Carnegie Mellon University} 
  \city{Pittsburgh}
  \country{United States}
}
\email{rbd@cs.cmu.edu}

\author{Kejun Zhang}
\authornote{Corresponding author}
\affiliation{%
  \institution{Zhejiang University}
  \city{Hangzhou}
  \country{China}
}
\affiliation{%
  \institution{Innovation Center of Yangtze River Delta, Zhejiang University}
  \city{Hangzhou}
  \country{China}
}
\email{zhangkejun@zju.edu.cn}


\renewcommand{\shortauthors}{Zihao Wang et al.}

\begin{abstract}
Singing accent research is underexplored compared to speech accent studies, primarily due to the scarcity of suitable datasets. Existing singing datasets often suffer from detail loss, frequently resulting from the vocal-instrumental separation process. Additionally, they often lack regional accent annotations.
To address this, we introduce the Multi-Accent Mandarin Dry-Vocal Singing Dataset (MADVSD). MADVSD comprises over 670 hours of dry vocal recordings from 4,206 native Mandarin speakers across nine distinct Chinese regions. In addition to each participant recording audio of three popular songs in their native accent, they also recorded phonetic exercises covering all Mandarin vowels and a full octave range.
We validated MADVSD through benchmark experiments in singing accent recognition, demonstrating its utility for evaluating state-of-the-art speech models in singing contexts. Furthermore, we explored dialectal influences on singing accent and analyzed the role of vowels in accentual variations, leveraging MADVSD's unique phonetic exercises.

\end{abstract}



\maketitle

\section{Introduction}

In recent years, speech and music models have developed rapidly~\cite{copet2024simple, dhariwal2020jukebox, 2022songdriverarxiv, wang2024remastarxiv, 10222365, wang2024muchinarxiv, 2024muditarxiv}. However, in the field of accent research, the singing domain significantly lags behind the speech domain. This is primarily due to the lack of high-quality accent datasets in the singing field~\cite{huang2023singing}. Existing large-scale singing datasets are mostly wet vocal tracks extracted from songs~\cite{bertin2011million}. Audio effects processing (such as reverb, echo, equalization, etc.) inevitably sacrifices original sound details. Furthermore, existing dry vocal singing datasets~\cite{Zhang2022neurips, Huang2021acm} generally lack regional accent labels.

To bridge this data gap and facilitate focused research on singing accent, we introduce the Multi-Accent Mandarin Dry Singing Vocal Dataset (MADVSD)\footnote{The dataset is available upon request for research purposes at \url{https://github.com/CarlWangChina/MADVSD}.}. MADVSD is a relatively large-scale, meticulously curated dataset comprising over 670 hours of dry vocal recordings from 4,206 native Mandarin speakers across diverse geographical regions of China. Collaborating with a nationwide organization, their employees across various provinces and cities were coordinated to record dry vocals at home using their mobile phones. These participants were recorded performing popular Mandarin songs and specifically designed phonetic exercises. These exercises encompass all Mandarin vowels and a full octave range of scales, providing granular phonetic and acoustic data for accent analysis. The dataset divides China into nine distinct accent regions, balancing regional specificity with sufficient speaker representation per region.

To demonstrate the utility of MADVSD, we conducted comprehensive experiments in the primary task of singing accent recognition. We benchmarked a range of state-of-the-art models, including audio pre-training models such as MelGAN~\cite{melgan}, VGGish~\cite{vgg}, YAMNet~\cite{yamnet}, Wav2Vec2.0~\cite{wav2vec2}, and HuBERT~\cite{hubert}, as well as models specialized for speech subdialect and accent recognition like KeSpeech~\cite{kespeech}, TeleSpeech-Pretrain-L\footnote{\url{https://github.com/Tele-AI/TeleSpeech-ASR}}, Qifusion~\cite{qifusion}, and DIMNet~\cite{dimnet}. Our results highlight the effectiveness of models like DIMNet in capturing singing accent characteristics.

Furthermore, through comparative experiments with the KeSpeech dataset~\cite{kespeech}, we investigated the correlation between dialectal influence and singing accent. Through the analysis of the extra phonetic exercises data, we also delved into the crucial vowel variations that contribute to accentual differences in singing.

Our contributions are summarized as follows:

1. Large-Scale Dataset: MADVSD is a new, large-scale dataset created to address the lack of singing accent data. It includes 12,618 dry vocal recordings of songs with regional accents, performed by 4,206 speakers from diverse regions in China. To our knowledge, it is the largest dry vocal singing dataset that includes regional accent labels.

2. Benchmark Validation: The dataset's effectiveness is validated through benchmark tests for the key task of Singing Accent Recognition. This test demonstrates MADVSD's utility in evaluating speech models applied to singing accent tasks.

3. Phonetic Resource for Research: MADVSD includes additional phonetic exercises with consistent text covering scales and vowels, recorded by all 4,206 speakers. This unique component provides a valuable resource for phonetic studies, specifically for investigating the influence of vowels on regional accents in singing.

\section{Background}

\subsection{Dialects and Accents}

In linguistics, dialects and accents are distinct yet related concepts. Dialects typically refer to language variations used within specific geographical regions. These variations can manifest significantly in vocabulary, grammar, and pronunciation, diverging from a standard language or other dialects~\cite{alharbi2021}.  For instance, Chinese dialects exhibit substantial differences from Mandarin Chinese in phonetics, lexicon, and syntax. These disparities can sometimes impede mutual comprehension between speakers from different dialect areas~\cite{norman2003chinese}.

Accents, conversely, generally denote pronunciation characteristics formed on the basis of a standard or common language, influenced by factors such as regional origin, social groups, or individual habits. Mandarin accents, for example, represent variations in pronunciation of Standard Mandarin influenced by regional differences and dialects~\cite{li2020tone}. While Mandarin with different accents may sound distinct, speakers with varying accents are generally able to understand each other, unless the pronunciation deviates severely from Standard Mandarin norms.

Dialectal studies have garnered considerable scholarly attention, as the diversity and complexity of dialects offer rich materials for linguistic research~\cite{list2015network}. In contrast, regional accent studies have been comparatively less explored, possibly due to the less salient nature of accentual variations compared to dialectal differences~\cite{guo1949relationship}.

\subsection{Dialect Datasets}

Dialect datasets have attracted substantial interest from researchers. Notable examples include KeSpeech~\cite{kespeech}, the Nexdata Chinese Dialect Dataset\footnote{\url{https://hf-mirror.com/datasets/Nexdata/chinese_dialect}}, and the Ten Thousand People Speaks Chinese Dialect Corpus\footnote{\url{https://hf-mirror.com/datasets/DataoceanAI/Ten_Thousand_People_Dialect_with_High_Quality_Labeling_Speech_Corpus}}. KeSpeech is a comprehensive dataset encompassing both Mandarin dialects and Standard Mandarin, featuring a total of 1542 hours of speech data from 27,237 speakers. This dataset encompasses eight distinct Mandarin dialect categories: Beijing, Southwestern, Zhongyuan, Northeastern, Lanyin, Jianghuai, Jilu, and Jiaoliao Mandarin. These dialects exhibit marked variations in vocabulary, grammar, and pronunciation compared to Standard Mandarin.
The Nexdata Chinese Dialect Dataset comprises 25,000 hours of speech data, including dialects such as Minnan, Cantonese, Sichuan, Henan, Northeastern, Shanghai, as well as Uyghur and Tibetan languages.
The Ten Thousand People Speaks Chinese Dialect Corpus is composed of 34,073 hours of speech data recorded by 29,954 speakers across 26 provinces in China. However, it is noteworthy that none of these datasets include unaccompanied singing data in dialects.

\subsection{Regional Accent Datasets}

In the realm of Mandarin regional accent datasets, AISHELL-1~\cite{aishell1} and DIDISPEECH~\cite{didispeech} datasets provide speaker labels indicating hometown or regional origin. The AISHELL-1 dataset encompasses recordings from 400 speakers from diverse regions of China, featuring a variety of Mandarin regional accents. Speakers originate from different accent regions, including Northern, Southern, and Cantonese-Guangxi-Minnan language areas. The DIDISPEECH dataset contains 800 hours of speech data from 6,000 speakers, providing label information on gender and regional distribution.

Furthermore, datasets like CommonVoice\footnote{\url{https://commonvoice.mozilla.org/zh-CN}}, CN-Celeb~\cite{cnceleb}, and Thchs-30~\cite{thchs} include Mandarin data with multiple regional accents, although the regional information provided is either incomplete or not comprehensive. The CommonVoice dataset, multilingual in nature, includes approximately 1,000 hours of Mandarin data contributed by over 40,000 contributors, with some metadata on accent, age, and gender. The CN-Celeb dataset comprises speech data from 1,000 speakers from various regions in China. The Thchs-30 dataset contains 30 hours of speech data from 50 speakers from diverse regions.

Nevertheless, it is important to note that these datasets also lack unaccompanied singing data featuring regional accents.

\subsection{Dry Singing Vocal Datasets}

Currently, commonly used open-source dry singing vocal datasets in academic research include: JSUT-Song~\cite{Sonobe2017arXiv}, which provides Japanese singing recordings; PJS~\cite{Koguchi2020JSJ}, a phoneme-balanced Japanese singing corpus recorded in a quiet environment; KiSing~\cite{Shi2022interspeech}, a singing voice corpus; Jvs Music~\cite{Tamaru2020arXiv}, featuring studio recordings from 100 Japanese singers; CSD~\cite{Choi2020ismir}, a children's song dataset performed by professional singers; Opencpop~\cite{Wang2022interspeech}, a high-quality Mandarin popular song corpus recorded in a studio; DSD100~\cite{Liutkus2017LVAICA}, which contains isolated dry vocal stems; Popcs~\cite{Liu2022aaai}, released as part of the DiffSinger model; M4Singer~\cite{Zhang2022neurips}, a multi-style, multi-singer professionally recorded Mandarin singing corpus; SingStyle111~\cite{dai2023singstyle111}, a professionally-recorded, multilingual, and multi-style dataset for singing style transfer; and OpenSinger~\cite{Huang2021acm}, a large-scale multi-singer singing voice corpus. However, to the best of our knowledge, none of these dry singing vocal datasets include specific labels for singers' regional accents.

\subsection{Recent Progress in AI Music Generation}

AI music generation is rapidly evolving from dataset construction to interactive, controllable systems for tasks like real-time accompaniment~\cite{2022songdriver} and emotion-based arrangement~\cite{remast}. A key challenge is aligning models with human intent, which has spurred work on new benchmarks~\cite{wang2024muchin} and user-aligned models~\cite{mudit-acl}. Concurrently, foundational models are tackling complex tasks such as full-song generation~\cite{yuan2025yue}, melody pre-training~\cite{wu2023melodyglm}, and zero-shot singing voice conversion~\cite{wang2024samoyearxiv}. This progress underscores the need for high-quality, annotated data, which our work provides for the under-explored domain of singing accents.

\section{Dataset Construction of MADVSD}

\subsection{Data Collection and Recording Setup}
In collaboration with the nationwide chain organization \textit{Mei KTV}, we coordinated with their employees across various provinces and cities to participate in the data collection. To capture authentic regional accents, participants recorded the dry vocals at home using their own mobile phones. They were instructed to record in a quiet indoor environment and to use pillows, quilts, or other soft materials to create a makeshift vocal booth to reduce ambient noise and reverberation.

\subsection{Participant Recruitment and Training}
The participants in this study were employees from various local branches of \textit{Mei KTV}. This approach ensured a wide geographical distribution of speakers. Through this collaboration, a cohort of 4,206 individuals was finalized. Participants were instructed to state their hometown at the beginning of each recording. This information was subsequently extracted automatically using Automatic Speech Recognition (ASR) and large language model-based information extraction techniques.

Initial training efforts explored having participants perform songs in their native dialects. However, this approach proved challenging for many, often resulting in performance disruptions due to laughter and pitch deviations, which led to recording failures. Consequently, we discontinued this method. For the formal data collection, we adopted a more pragmatic strategy, instructing participants to sing in Mandarin while retaining their regional accents.

\subsection{Dataset Composition and Content}
The finalized MADVSD dataset comprises 670 hours of recordings from 4,206 speakers, encompassing 12,618 full-length a cappella song performances and 4,206 segments of vowel vocalizations.

While participants were granted autonomy in song selection, the dataset's musical repertoire is primarily confined to Mandarin pop music, encompassing genres such as ballad pop, rock-infused pop, and folk ballads, reflective of participant musical backgrounds. Notably, genres considered either niche or technically demanding in vocal performance, such as rap, bel canto, traditional folk music, operatic arias, and Peking opera, are absent from this dataset.

\begin{figure}[htbp]
    \centering
    \includegraphics[width=1\linewidth]{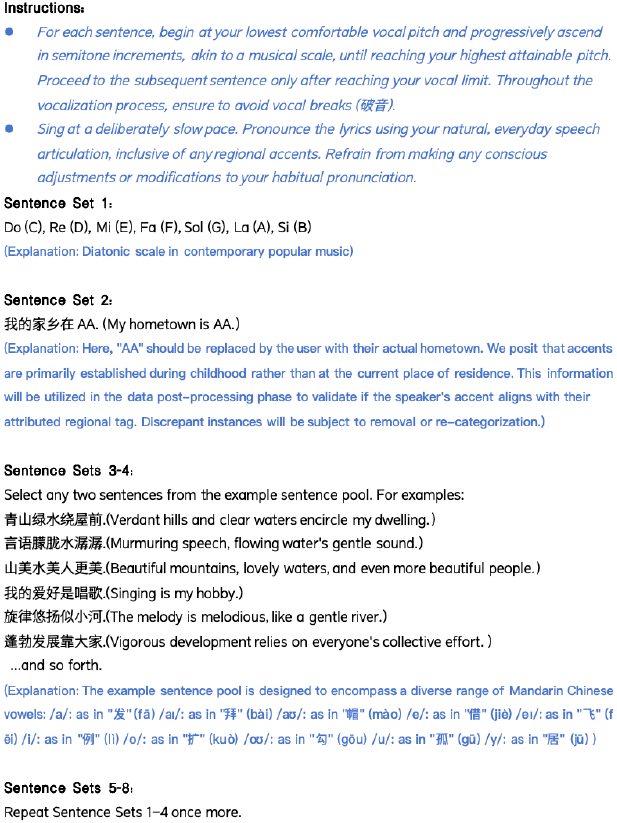}
    \caption{The content of the Vowel Phonetic Exercises Protocol.}
    \Description{A text document listing the instructions for a phonetic exercise protocol. The protocol is divided into several sentence sets. Sentence Set 1 consists of the notes of a diatonic musical scale (Do, Re, Mi). Sentence Set 2 is a template sentence in Chinese where the speaker inserts their hometown. Sentence Sets 3 and 4 are various example sentences in Mandarin designed to cover a wide range of Chinese vowels. The final instruction is to repeat the first four sets.}
    \label{fig:vowel}
\end{figure}

Participants contributed to a corpus of 4,206 standardized phonetic exercises recordings. The primary motivation for including this component was to facilitate phonological and linguistic investigations into the underlying principles of regional accent variations in singing. These vocalizations were designed to encompass the full spectrum of Chinese vowels. The content of these \textbf{phonetic exercises} is depicted in Figure~\ref{fig:vowel}.

\subsection{Formal Recording Procedures and Guidelines}
For each participant, the recording process commenced with a selection of three songs of their choice, totaling approximately 8-12 minutes of performance per participant. Participants were instructed to monitor musical accompaniment via headphones while recording their a cappella vocals using their mobile phones. Rigorous measures were taken during recording to prevent instances of microphone plosives and audio distortion. Participants were further instructed to ensure the acoustic clarity of each word and note.

Following song recordings, all 4,206 participants were also tasked with recording a set of phonetic exercises, with each set of exercises lasting about 1-2 minutes. These vocalizations, combined with the song recordings constitute the two types of dry vocals collected from each singer.

\subsection{Accent Region Demarcation}
Upon completion of data collection, the subsequent phase involved the demarcation of accent regions. This process required balancing the granularity of accent types with sufficient speaker representation per category to ensure model training efficacy. While our recordings spanned every province and city, we demarcated the accent regions by prioritizing adequate speaker counts for each category, which resulted in a manageable set of nine accent types for the identification task.

This methodological choice led to a specific geographical pattern: smaller, more densely populated accent regions were defined in eastern coastal areas with a higher speaker count, whereas geographically larger regions correspond to the central and western areas with a lower speaker count. We acknowledge that our data collection, conducted in collaboration with \textit{Mei KTV}, was constrained by their branch locations, which hindered data acquisition from all provinces. Future endeavors will explore alternative strategies to incorporate data from these currently uncovered regions. For readers interested in a visual reference, our project repository features a map that color-codes the geography into nine distinct regions to visualize the speaker distribution\footnote{\url{https://github.com/CarlWangChina/MADVSD}}.

The final speaker distribution across the nine demarcated accent regions is detailed in Table~\ref{tab:accent_regions}.

\begin{table}[htbp]
    \centering
    \caption{Speaker distribution across accent regions.}
    \label{tab:accent_regions}
    \small
    \setlength\tabcolsep{4pt}
    \begin{tabularx}{\columnwidth}{lXr}
        \toprule
        \textbf{Abbre.} & \textbf{Region Full Name} & \textbf{Speaker Count} \\
        \midrule
        SZR & Shanghai-Zhejiang Region & 782 \\
        JR & Jiangsu Region & 290 \\
        SSHR & Shanxi-Shandong-Henan Region & 362 \\
        BMLR & Beijing-InnerMongolia-Liaoning Region & 614 \\
        SGXR & Shaanxi-Gansu-Xinjiang Region & 400 \\
        FR & Fujian Region & 455 \\
        HHAR & Hunan-Hubei-Anhui Region & 438 \\
        GGR & Guangdong-Guangxi Region & 484 \\
        YGSR & Yunnan-Guizhou-Sichuan Region & 381 \\
        \midrule
        \textbf{Total} & & \textbf{4,206} \\
        \bottomrule
    \end{tabularx}
\end{table}

\subsection{Recorded Data Post-Processing}
The use of non-uniform recording devices presented a challenge for data consistency in model training. To mitigate this, we employed tools like LALAL.AI and UVR5 to standardize the quality and acoustic characteristics of the audio data. Initially, recordings were processed to separate left and right audio channels. The channel exhibiting greater signal amplitude was then selected and converted to mono format to streamline subsequent AI model processing. Subsequently, noise reduction and dereverberation were performed using tools such as LALALAI and UVR5 to mitigate environmental artifacts. Finally, frequencies below 80Hz were filtered out to minimize the potential adverse effects of low-frequency noise on subsequent model training.

\subsection{Data Availability and Privacy Protocol}
Each speaker is annotated with metadata including accent region, province, and city of origin. Given the sensitive nature of the audio data, which contains personally identifiable information such as hometown, name references, and unique vocal characteristics, the dataset is subject to privacy protocols. 
Researchers should complete the application form on the project's repository homepage and email it to the designated repository administrator. Upon joint review and approval by our team and the collaborating organization, an agreement will be executed to authorize data access for research purposes exclusively.

\section{Experiment}

\subsection{Singing Accent Recognition}
\subsubsection{Task Overview}
Analogous to speech accent recognition, singing accent recognition primarily involves inputting a user's a cappella singing to determine their accent, origin, and potential hometown.

\subsubsection{Models Under Evaluation}
In this task, we evaluated two categories of models:

(1) \textbf{Audio Pre-training Models}: For these models, we extracted embeddings and fed them into a Multilayer Perceptron (MLP). The MLP was fine-tuned to output accent classification results.
MelGAN~\cite{melgan}: While primarily designed for decoding Mel spectrograms into waveforms, its encoder can also be utilized for feature extraction.
VGGish~\cite{vgg}: Developed by Google and based on the classic VGG network, VGGish is suitable for environmental sound classification and audio emotion recognition tasks.
YAMNet~\cite{yamnet}: A lightweight audio embedding model designed for resource-constrained scenarios, YAMNet is applicable for real-time audio processing.
Wav2Vec2.0~\cite{wav2vec2}: Developed by Facebook AI, Wav2Vec2.0 focuses on self-supervised learning to acquire rich speech representations from raw audio.
HuBERT~\cite{hubert}: Also from Facebook AI, HuBERT iteratively refines speech representations through the use of discrete clustering.

(2) \textbf{Speech Subdialect/Accent Recognition Models}: These models can be directly transferred to the singing accent recognition task.
KeSpeech~\cite{kespeech}: The baseline model from the KeSpeech dataset, designed for speech subdialect identification.
TeleSpeech-Pretrain-L\footnote{\url{https://github.com/Tele-AI/TeleSpeech-ASR}}: The large configuration of TeleSpeech-ASR 1.0, an automatic speech recognition model developed by Tele-AI and pre-trained on a substantial speech dataset.
Qifusion~\cite{qifusion}: Qifusion-Net is a Layer-Adapted Fusion (LAF) model designed for end-to-end multi-accent speech recognition, primarily based on the KeSpeech dataset.
DIMNet~\cite{dimnet}: Based on KeSpeech, DIMNet is designed to handle multi-accent scenarios by explicitly modeling the interaction between Automatic Speech Recognition (ASR) and Accent Recognition (AR) tasks.

\subsubsection{Evaluation Metrics}

\textbf{MR-ACC: Multi-Region Accent Classification Accuracy}. This metric evaluates the capability of multi-region accent classification, specifically extracting samples from 9 distinct regions and identifying their accent type. This task is relatively complex due to the differentiation among multiple regions.
\textbf{JR-AD, SSHR-AD, SZR-AD, etc. (Nine metrics in total)}. Each of these metrics focuses on accent recognition within a specific region, determining whether an audio sample belongs to that region (Yes/No). These are simpler binary classification tasks, contrasting a single region against all others. For example, SZR-AD stands for Shanghai-Zhejiang Region Accent Detection. The other metrics follow a similar naming convention.

\subsubsection{Experimental Results}

\begin{figure}[htbp]
    \includegraphics[width=1\linewidth]{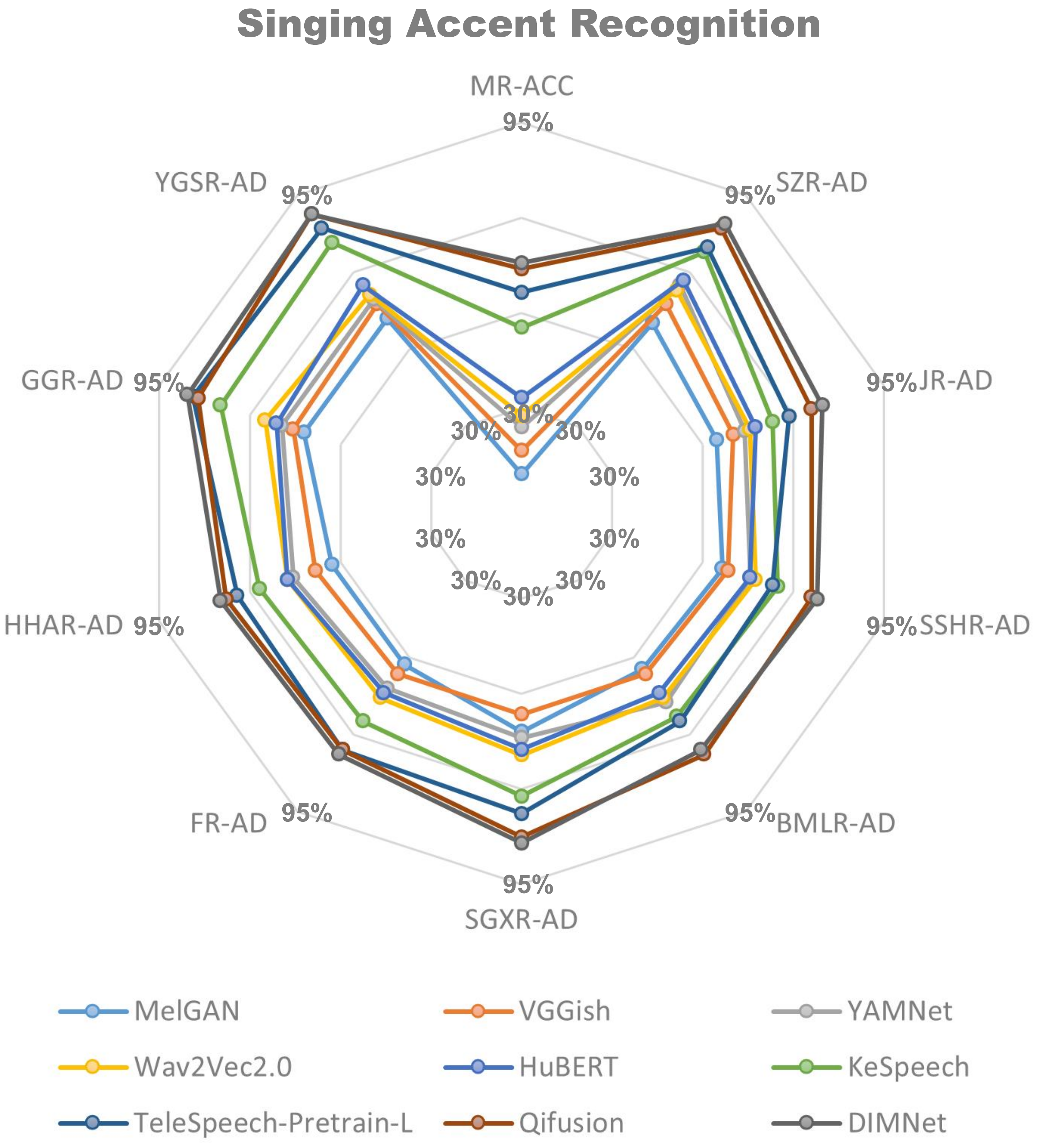}
    \caption{Performance of Accent Recognition Models}
    \Description{A radar chart comparing the performance of nine different models on ten accent recognition tasks. The models, including DIMNet and Qifusion, are represented by different colored lines. The chart's axes represent metrics like MR-ACC and region-specific accent detection such as SZR-AD and YGSR-AD. The lines for DIMNet and Qifusion form the outermost layers of the chart, indicating they achieved the highest performance across most tasks.}
    \label{fig-accent-recog}
\end{figure}

The experimental results, as depicted in Figure~\ref{fig-accent-recog}, indicate that DIMNet achieved the best performance across most metrics, with the exception of BMLR-AD and YGSR-AD, where Qifusion slightly outperformed it. This suggests that DIMNet's architecture maintains strong generalization capabilities in singing accent recognition.

The Multi-Region Accent Classification Accuracy is notably lower than the Accent Detection accuracy for individual regions. This implies that binary classification tasks (confirming whether an accent belongs to a specific region) are easier for models to train effectively, whereas distinguishing among 9 different accent types poses a greater challenge.

Furthermore, the Accent Detection (AD) scores for GGR-AD, YGSR-AD, SZR-AD, and SGXR-AD are higher than those of other regions. This suggests that these four regions possess more distinctive regional characteristics and exhibit greater divergence from standard Mandarin, making their features relatively easier for models to capture.

The scores of speech subdialect/accent recognition models generally surpass those of audio pre-training models. However, upon transferring speech subdialect/accent recognition models to this task, their performance is, in fact, lower than their original performance on speech accent recognition tasks using similar metrics. This performance degradation may be attributed to the data discrepancy between singing and speech, as well as the inherent differences between accent Mandarin and dialects.

\subsection{Correlation Analysis between Dialect and Accent}

Previous research has indicated that accents are variations in pronunciation of standard Mandarin influenced by dialects~\cite{li2020tone}. In this experiment, we aim to validate this claim empirically.

We referenced the model architecture for subdialect recognition in KeSpeech and retrained it on MADVSD as a control group. We then compared its performance with the KeSpeech subdialect recognition baseline and the results after further training the baseline on MADVSD. By comparing the performance differences among these three configurations, we analyzed whether dialect data contributes to accent recognition. To ensure a fair and direct comparison, all models in this experiment were evaluated on a consistent, held-out test set partitioned from the MADVSD dataset, using the MR-ACC metric.

\begin{table}[htbp]
    \centering
    \caption{Performance of the KeSpeech subdialect model on the MADVSD test set after training on various dataset combinations.}
    \label{tab:dialect_accent_corr}
    \begin{tabular}{lccc}
        \toprule
        & KeSpeech & MADVSD & KeSpeech + MADVSD \\
        \midrule
        MR-ACC & 51.3\% & 56.8\% & 60.1\% \\
        \bottomrule
    \end{tabular}
\end{table}

The results, as shown in Table~\ref{tab:dialect_accent_corr}, indicate that our dataset performs better than KeSpeech in the singing accent recognition task, demonstrating the unique value of MADVSD. 

Furthermore, KeSpeech+MADVSD achieved the highest score, suggesting that the dialect speech data included in the pre-trained model aids in enhancing the performance of singing accent recognition. This further substantiates the notion that regional accents are influenced by local dialects.

It is important to note that the KeSpeech performance presented in the preceding radar chart reflects the scores of KeSpeech+MADVSD, rather than its original performance.

\subsection{Key Vowel Analysis in Singing Accent}

Standard Mandarin Chinese comprises approximately 32 phonemes, including 10 vowel phonemes and 22 consonant phonemes. These are critical components in shaping the characteristics of dialects and accents.

We aim to leverage annotated datasets to analyze and investigate the underlying principles of accent differences in singing across various regions, specifically to identify which vowel variations are more influential. In this study, we focus on YGSR and SSHR regions as illustrative examples.

\subsubsection{Analysis Metrics}
\textbf{Vowel Analysis Metric (VAM)}: This metric quantifies the difference in vowel pronunciation between two accents (YGSR and SSHR) by using the cosine similarity of embeddings from a state-of-the-art (SOTA) accent recognition model (DIMNet). Lower cosine similarity suggests a greater difference in vowel pronunciation. We first use the automated tool Montreal Forced Aligner (MFA)~\cite{mfa} to perform a preliminary phoneme alignment, in order to crop and extract the corresponding vowel audio segments.

\subsubsection{Analysis Results}

\begin{figure}[htbp]
	\includegraphics[width=1\linewidth]{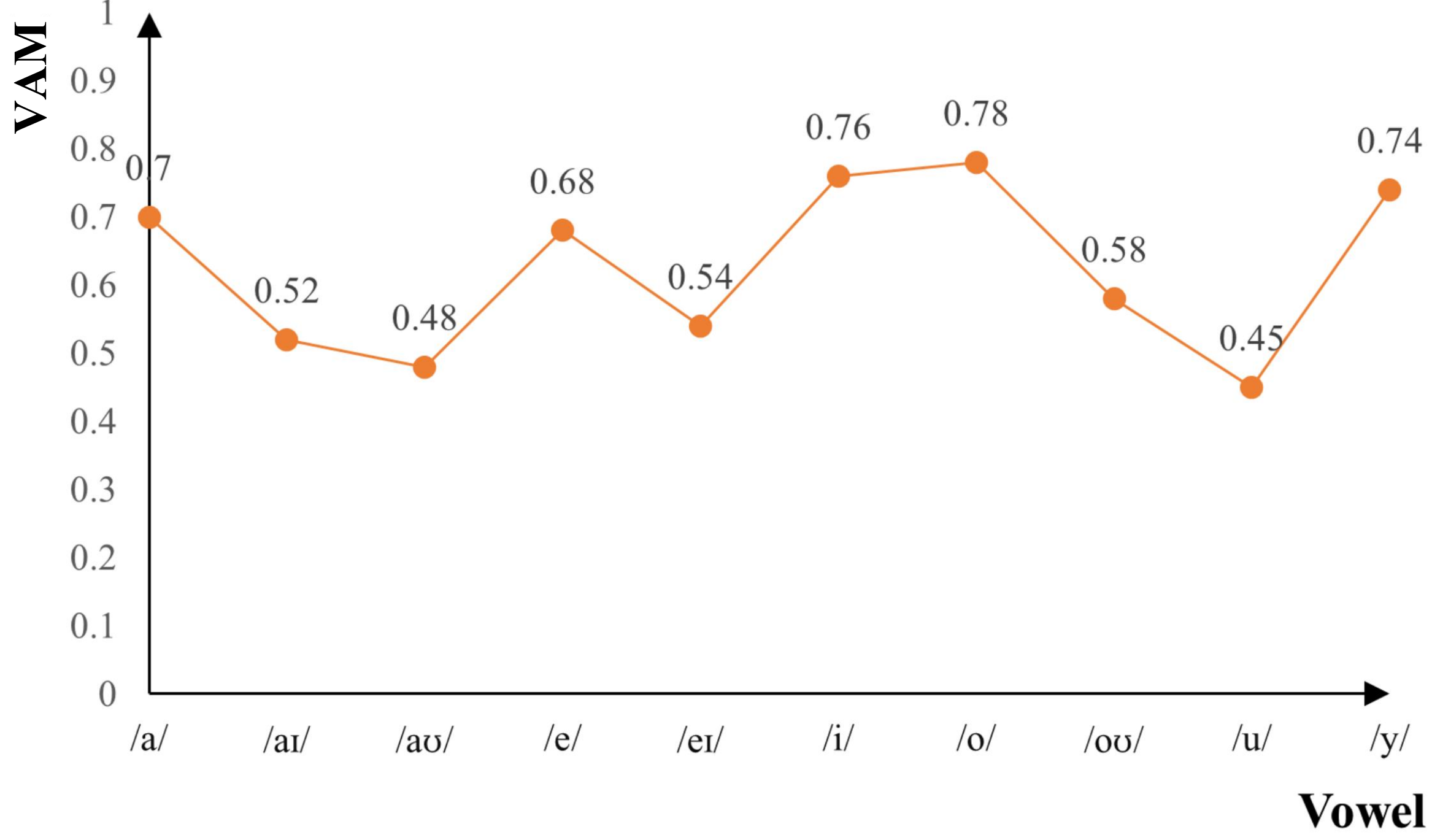}
    \caption{Vowel Analysis for YGSR and SSHR Accents.}
    \Description{A line graph plotting a metric labeled SAM, ranging from 0 to 1, for ten different Mandarin vowels shown on the x-axis. The vowels include /a/, /aɪ/, /aʊ/, /e/, /eɪ/, /i/, /o/, /oʊ/, /u/, and /y/. The line shows significant variation across vowels, with the metric's value peaking at 0.78 for the vowel /o/ and reaching its lowest point at 0.45 for the vowel /u/.}
	\label{vowel-result}
\end{figure}

The analysis results, as depicted in Figure \ref{vowel-result}, present the Vowel Analysis Metric (VAM) scores that quantify the pronunciation differences for ten Mandarin vowels between the YGSR and SSHR accents. According to the metric's definition, a lower cosine similarity score indicates a more significant pronunciation difference. The data reveals that the most substantial accentual divergences are found in the vowels /u/, /a\textipa{U}/, and /a\textipa{I}/. Specifically, the vowel /u/ exhibits the greatest variation with the lowest VAM score of $0.45$, followed by /a\textipa{U}/ ($0.48$) and /a\textipa{I}/ ($0.52$). Conversely, the simple vowels /o/, /i/, and /y/ are the most similarly pronounced, with /o/ yielding the highest VAM score of $0.78$, indicating the least accentual difference. These findings suggest that complex diphthongs and the high back rounded vowel /u/ are the key phonetic markers that most significantly distinguish the YGSR and SSHR regional accents in the context of singing.

\section{Conclusion and Future Work}

This paper presents MADVSD, a novel and substantial dataset of Mandarin singing with diverse regional accents. We detail its establishment procedures, conduct comprehensive baseline experiments for accent recognition, and perform insightful phonetic analysis, offering preliminary findings on the characteristics of singing accents based on this new dataset.

For future work, we plan to extend this research in several promising directions. Firstly, we will venture into the field of \textbf{Singing Accent Conversion}, with the goal of developing models that can transform a vocal performance from one regional accent to another while preserving the singer's identity and musical expression. Secondly, we intend to explore the \textbf{influence of accents on singing aesthetics}. This line of inquiry will involve a more granular analysis of the subtle differences among accents in terms of pitch accuracy, rhythmic precision, and phonetic articulation, aiming to understand how these variations are perceived by listeners and contribute to the overall artistic impression of a performance.

\section{Limitation}
While this paper introduces the MADVSD dataset for Singing Accent Recognition and provides initial analyses, we acknowledge several limitations:

\textbf{Limited Music Genre Coverage:} The dataset primarily focuses on Mandarin pop music (e.g., ballads, rock, folk), lacking diversity in genres like rap, bel canto, opera, and traditional Chinese opera. This genre bias may impede model generalization. Future work should broaden genre scope to enhance model robustness across diverse musical styles.

\textbf{Restricted Geographical Coverage:} Geographical coverage is restricted due to data collection being limited to partner institution locations. This results in missing accent data from certain provinces, potentially affecting the dataset's representativeness of all Chinese regional accents. Future efforts will explore broader data collection via more collaborations or online methods to achieve comprehensive geographical coverage and improve regional representativeness.

\textbf{Broad Accent Region Divisions:} Accent region divisions are broad, with a nine-region classification where some regions (e.g., SGXR) are geographically vast while others are smaller. This broad classification may not fully capture subtle accent variations. Future work, with increased data, should adopt a more granular division, potentially at the province or city level, for more accurate accent representation.

\textbf{Potential Influence of Participants' Dialect Backgrounds:} Participant dialect backgrounds may influence accent purity. While participants sang Mandarin with accents, varying dialect backgrounds and Mandarin proficiency may introduce variability, impacting accent research. Future datasets should include participant dialect background information or stricter screening to control this influence and improve dataset homogeneity.

\begin{acks}
This work was supported by the National Natural Science Foundation of China (No.62272409).
\end{acks}

\bibliographystyle{ACM-Reference-Format}
\balance
\bibliography{sample-base}

\end{document}